\documentstyle[11pt]{article}
\begin{document}
\thispagestyle{empty}
\baselineskip 18pt
\rightline{KIAS-P02022}
\rightline{UOSTP-02004}
\rightline{{\tt hep-th/0206185}}
%\large
\

\def\tr{{\rm tr}\,}
\def\Tr{{\rm Tr}\,}
\newcommand{\beq}{\begin{equation}}
\newcommand{\eeq}{\end{equation}} \newcommand{\beqn}{\begin{eqnarray}}
\newcommand{\eeqn}{\end{eqnarray}} \newcommand{\bde}{{\bf e}}
\newcommand{\balpha}{{\mbox{\boldmath $\alpha$}}}
\newcommand{\bsalpha}{{\mbox{\boldmath $\scriptstyle\alpha$}}}
\newcommand{\betabf}{{\mbox{\boldmath $\beta$}}}
\newcommand{\bgamma}{{\mbox{\boldmath $\gamma$}}}
\newcommand{\bbeta}{{\mbox{\boldmath $\scriptstyle\beta$}}}
\newcommand{\btau}{{\mbox{\boldmath $\tau$}}}
\newcommand{\lambdabf}{{\mbox{\boldmath $\lambda$}}}
\newcommand{\bphi}{{\mbox{\boldmath $\phi$}}}
\newcommand{\bslambda}{{\mbox{\boldmath $\scriptstyle\lambda$}}}
\newcommand{\ggg}{{\boldmath \gamma}} \newcommand{\ddd}{{\boldmath
\delta}} \newcommand{\mmm}{{\boldmath \mu}}
\newcommand{\nnn}{{\boldmath \nu}}
\newcommand{\diag}{{\rm diag}}
\newcommand{\bra}{\langle}
\newcommand{\ket}{\rangle}
\newcommand{\sn}{{\rm sn}}
\newcommand{\cn}{{\rm cn}}
\newcommand{\dn}{{\rm dn}}
\newcommand{\tA}{{\tilde{A}}}
\newcommand{\tphi}{{\tilde\phi}}
\newcommand{\bpartial}{{\bar\partial}}
\newcommand{\br}{{{\bf r}}}
\newcommand{\bx}{{{\bf x}}}
\newcommand{\bk}{{{\bf k}}}
\newcommand{\bq}{{{\bf q}}}
\newcommand{\bQ}{{{\bf Q}}}
\newcommand{\bp}{{{\bf p}}}
\newcommand{\bP}{{{\bf P}}}
\newcommand{\thet}{{{\theta}}}
\newcommand{\tauu}{{{\tau}}}
\renewcommand{\thefootnote}{\fnsymbol{footnote}}
\newcommand{\E}{{{\cal E}}}
\newcommand{\N}{{\scriptscriptstyle N}}

\vskip 0cm
\centerline{ \LARGE \bf Supertubes 
connecting  D4 branes }

\vskip .2cm

\vskip 1.2cm
\centerline{ \large Dongsu Bak$\,^a$\footnote{Electronic Mail:
dsbak@mach.uos.ac.kr}  and 
Kimyeong Lee$\,^{b}$\footnote{Electronic Mail:
klee@kias.re.kr} 
}

\vskip 3mm
\centerline{\it $^a$Physics Department, University of Seoul, Seoul
130-743, Korea}
\vskip 3mm
\centerline{ \it $^b$ School of Physics, Korea Institute for Advanced
Study}
\centerline{ \it 207-43, Cheongryangri-Dong, Dongdaemun-Gu, Seoul
130-012, Korea
}
\vskip 1.2cm

%\vskip 1.2cm

\begin{quote}
{We find and explore a class of dyonic instanton solutions which can
be identified as the supertubes %and super D2-$\overline{\rm D2}$ 
connecting two D4 branes. They
correspond to a single monopole string and a pair of
monopole-antimonopole strings from the worldvolume 
view point of D4 branes.} 
\end{quote}
%\centerline{\today}

%\pacs{14.80.Hv,11.27.+d,14.40.-n}

\newpage
\baselineskip 18pt
\section{Introduction}

Recently there has been a considerable study of the so called
supertube solutions. Originally Mateos and Townsend constructed
supersymmetric configurations of D2-D0-F1 branes where D2 brane forms
a single cylinder with D-particles and F-strings melted in\cite{mateos}.
%and D0 branes lie on D2 and F1 string lies along the
%cylinder, say, along $x^9$\cite{mateos}. 
The corresponding configurations in the
matrix model have been found also\cite{klee,swkim}.  As a 
limit of the elliptic deformation of circular tubes, 
flat supersymmetric brane antibrane configurations are 
realized\cite{karch} and the wouldbe tachyons present in the 
ordinary brane-antibrane system become massless exhibiting the 
stability of the system\cite{ohta}.   
In fact the cross sectional shape of the tubes is shown to be
%can take any shape 
arbitrary
in the transverse eight dimensions\cite{ng}. 
%Any number of
%parallel supertubes would be also supersymmetric.  
More recently, the
DBI action has been explored to find the supersymmetric configuration
for D2 supertubes ending on a D4 brane\cite{kruczenski}. 
The higher dimensional analogue
has also been  studied\cite{kruczenski,jabbari,lugo} 
and there appeared other related 
works\cite{cho,emparan,townsend3,ohta2}.

In this note we are interested in tubular D2 connecting two D4 branes
separated, say, in $x^9$ direction. As argued in
Ref.~\cite{kruczenski}, all the constituents of supertubes, i.e.  D0,
F1 and D2, may end on D4 branes and, hence, such configuration should
be consistent with supersymmetries. Viewed from the worldvolume of D4,
D0 appears as an instanton generating selfdual field strength and
F-string extended along $x^9$ works as a localized electric
source. One direction of D2 is extended in $x^9$ and the other spatial
direction will in general form a curve in D4 worldvolume. Since this
curve is magnetically charged, one may call it as a magnetic string
from the view point of the D4 worldvolume. The instantons preserve
only half of sixteen supersymmetries and F1/D2 breaks further half, so
only 1/8 of 32 supersymmetries remain in general.

In this U(2) 
context of two D4 branes, one is ultimately interested in magnetic 
strings where the shape of  curve on D4 is arbitrary, 
which correspond to tubes of arbitrary cross 
sections. However such general configurations seem involved for 
reasons described below. Therefore we shall focus on the monopole and 
antimonopole
strings corresponding to  supersymmetric 
flat  D2-$\overline{\rm D2}$
connecting two D4 branes. 
This supersymmetric configuration 
%monopole antimonopole system  
is composed of a monopole string 
and an antimonopole
string.  

Simpler situation arises if one considers just one flat D2 (or a
$\overline{\rm D2}$) suspended between D4 branes, in which D0 and
fundamental strings are melted in.  As demonstrated below, such single
D2 preserves actually 8 real supersymmetries becoming 1/4 BPS.  When
one brings such charged monopole strings and an antimonopole string
together, one might naively expect that they cannot be BPS or static
due to attraction between opposite charges.  However this is not the
case as in the supersymmetric brane antibrane
systems\cite{karch,ohta}.  In spite of the presence of D4, still this
monopole antimonopole strings can be static preserving four real
supersymmetries\cite{kruczenski,jabbari}.  
Below we shall construct the super
D2-anti D2 configuration in the context of D4 whose worldvolume
fluctuations are described by super Yang-Mills theories.

For later comparison, we first briefly review here the key
characteristics of supersymmetric tubular D2 branes using the
Born-Infeld theory. One could consider the case of an arbitrary cross
section with most general B-field.  But such generality won't be
necessary because we are mainly concerned the flat D2 or
anti-D2. Furthermore for the fully local characteristics of tubular
branes, considering flat one suffices. The tubular configuration has a
translational symmetry in one direction which we shall take $x^9=z$
and the other spatial direction as $x^4=\sigma$ where $\sigma$ and $z$
will be used for the worldvolume coordinates. Turning on $E=F_{0z}$
and $B=F_{\sigma z}$, the Born-Infeld action becomes
\beq
{\cal L}= -\sqrt{-\det(g+F)}=-\sqrt{1-E^2+B^2}
\eeq
The displacement  $\Pi$ conjugated to $E$ becomes
\beq
\Pi={\partial {\cal L}\over  \partial E}={E\over
\sqrt{1-E^2+B^2}}\,. 
\eeq 
Below we shall consider   $E>0$ case without loss of 
generality.
The Hamiltonian 
may be arranged  as
\beq
{\cal H}=\sqrt{(1+B^2)(1+\Pi^2)}= 
\sqrt{(1-|B|\Pi)^2 +(|B|+\Pi)^2}\,\,\,\ge  |B|+\Pi\,.
\eeq
The BPS equation 
\beq
|B|\Pi=1
\eeq 
leads to the condition $E=1$. The Gauss law $\partial_z \Pi=0$
then implies that $B$ should be independent of $z$ but an 
arbitrary function of $\sigma$, i.e. $B=B(\sigma)$.  
The  $B$ field
 describes the spatial 
density of D0 branes melted into D2 because 
${1\over 2\pi}\int\! dz d\sigma\, B$ 
counts the total
 number of D0 branes while $\Pi$ is the linear density of
F-string extended in $z$ direction as the electric 
flux $\int d\sigma\, \Pi $ 
counts the total number of string charges. Further there is nonvanishing 
momentum density produced by the field. By evaluating Poynting 
vector, one finds that  the field 
momentum density %produced by the fields
is nonvanishing with  ${\cal P}_\sigma=1$. The key local characteristics
of supertubes 
are summarized as follows. 1) The F-string density $\Pi$ is related to
D0 density by $\Pi=|B|^{-1}$; as a result, the field momentum density 
%produced by the field 
satisfies ${\cal P}_\sigma=1$.
2) E=1. 3) ${\cal H}_{D2}= 
|B|+|\Pi|$.

In the discussion of the monopole strings, we shall see that the above
key structures are precisely reproduced.  Since the Born-Infeld
analysis involves highly nonlinear terms, why the super Yang-Mills
theories that are at most quadratic in their field strengths, have a
capacity to do so is not clear.  But note that all the above four
characteristics are faithfully reproduced in the matrix model
description\cite{klee,swkim,karch,jabbari}.  The matrix model is closely
related to the (noncommutative) Yang-Mills theories and this could
partially explain the reasons.

\section{Field theory set-up}

We are interested in the five dimensional super Yang-Mills theory with a
single adjoint scalar field $\phi= X^9$ turned on. The relevant  
bosonic part 
of Lagrangian is
\beq
{\cal L } = -\frac{1}{4g^2}  \Tr\left( F_{MN}F^{ MN} + 2 D_M\phi
D^\mu \phi \right)
\eeq
where $F_{MN}= \partial_M A_N - \partial_N A_M -i[A_M,A_N]$. We use
 the notation $A_M=A_M^a T^a$ where $\Tr T^a T^b = \delta^{ab}/2$
and $T^a= {\sigma^a/2}$ for the SU(2) case.
The energy functional can be bounded\footnote{$i,j,k=1,2,3$ 
and $\mu,\nu=1,2,3,4$.}
\beq
{\cal E} = \frac{1}{2 g^2}\int d^4x  \Tr \left\{  (F_{0\mu} \mp D_\mu
\phi)^2 + (B_i - F_{i4})^2  + (D_0\phi)^2 \right\} + 
\frac{4\pi^2}{g^2} \nu \pm {\cal Q}^E
\eeq
where $B_i= \epsilon_{ijk} F_{jk}/2$. 
\beqn
&& \nu =  \frac{1}{4\pi^2} \int d^4x \Tr B_i F_{i4} \\
&& {\cal Q}^E = \frac{1}{g^2} \int d^4x \partial_\mu  \Tr (F_{0\mu}\phi) \,.
\eeqn
The selfdual equation for the dyonic instanton is
\beqn
&& D_0\phi=0 \\
&& B_i = F_{i4} \label{bps1} \\
&& F_{\mu0} \mp D_\mu \phi=0 
\label{bps2}
\eeqn
The above equations and the  Gauss law $D_\mu F_{\mu 0} +i[\phi,
D_0\phi]=0$ imply 
\beq
D_\mu^2 \phi=0 
\label{laplace}
\eeq
The selfdual equation for the instanton alone preserves 8 real 
supersymmetries. The presence of the electric and the scalar components
breaks further half. Thus configurations satisfying the above BPS 
equations in general preserve 4 real supersymmetries.

While we can obtain a field configuration with $D_0\phi\ne 0$ by the
Lorentz boost of the above configurations, we will restrict our
discussion to the configurations such that $D_0\phi=0$.

In the SU(2) super Yang-Mills theories of two D4 branes, solutions of
the selfdual equation as well as $E_\mu= D_\mu \phi$ have already
appeared in the literature\cite{lambert,zamaklar,eyras}. 
(See  Ref.~\cite{sang} 
for a more recent
discussion on dyonic calorons.) They involve
instantons (D0's) as well as electric charges of fundamental strings
stretched to the $x^9$ direction. The separation $h$ of the two D4
branes is described by the vacuum expectation value of scalar field
$\bra\phi\ket={\sigma_3\over 2}h\,$.  These dyonic instantons solve
 precisely the same BPS equations for the supertubes connecting
separated D4 branes.  The $N$ dyonic instanton solution in the 't Hooft
ansatz reads
\beqn && A_\mu
={\sigma_a\over 2} \bar\eta^{a\mu\nu} \partial_\nu G\,,\nonumber\\ &&
\phi=X^9= {\sigma_3\over 2} {h\over G} \eeqn 
where
$\bar\eta^{a\mu\nu}$ is the antiselfdual 't Hooft tensor. 
%and $a$ is
%the vacuum expectation value of $X^9$ corresponding to the separation
%of two D4 in $x^9$ direction. 
The function $G$ is given by 
\beq 
 G =1
+\sum^N_{n=1} {\rho^2_n\over |{\bf x}-{\bf x}_n|^2}\,.  
\eeq 
where $\rho_n$ is for the size and $x^\mu_n$ represents the position
of $n$-th instanton.  This SU(2) configuration certainly carries
electric charges and the number of D0 is $N$. However the
configurations do not carry magnetic string charges. Since
$F_{\mu\nu}\sim 1/r^4$, the system carries at most dipole moment of
monopole charges. Thus, for example, one monopole string discussed
below does not follow straightforwardly by arranging them over a line
uniformly. Moreover considering $N=1$ dyonic instanton, %the shape of
the throat connecting two D2 branes has a topology $R\times S^3$,
which is not the geometry of tubular D2 branes.

There is well known way to get ordinary monopole solution independent
of $x^4$ from above instanton solution\cite{harry}. %, which is
%independent of $x^4$.  
First arrange all the instanton over $x^4$ axis
with equal spacing and make all the size equal i.e. $\rho_n=\rho$.  In
the limit where the size parameter goes to infinity, the above
solution $A_\mu$ becomes, up to a gauge transformation, one BPS
monopole solution independent of $x^4$, where $A_4$ plays
the role of scalar\cite{harry}. 
In this limit, $\phi\,\,\, (\,=X^9)$ vanishes and the
limiting solution does not correspond to a D2 (between D4 branes)
presented below. As we shall see below the limit where $\rho$ goes to
infinity corresponds to sending massless antimonopole to infinity. In
doing so, the magnetic charge as well as the electric charge become
different, so there is a change in their main physical content.

In short, what we like to argue here is that the above collection of
the dyonic instanton does not carry D2 charges.  Further they do not
satisfy the key characteristics of tubular D2 branes. For example if
one arranges them on a straight line on D4 uniformly with equal spacing
$\Delta$, then the instanton number density is proportional
$1/\Delta$. The total electric charge is again proportional to
$1/\Delta$. This kind of configuration then does not satisfy the
relation, say, $\Pi B=1$ in general.  The momentum density ${\cal
P}=1$ is not produced either.  Instead they carry only nonvanishing
angular momentum.

The solutions given by the 't Hooft ansatz could be interpreted as the
collapsed supertubes.  The configurations beyond the 't Hooft ansatz
may have a tube structure but it remains to be seen whether this is
true.  For this reason, we shall construct the flat charged D2 or
super D2-$\overline{\rm D2}$ connecting D4 branes, which are simpler
examples of supertubes.

\section{Single monopole string}

Let us  start with the  magnetic monopole-like string which has
uniform instanton density along the $x^4$ direction. The configuration
is given by $A_i^a= V_i^a $ and  $A_4^a= V_4^a$ where 
\beqn
&& V_i({\bf r},u) \equiv \epsilon^{aij} \frac{\sigma^a}{2}\frac{\hat{r}^j}{r} 
\left( 1- \frac{u r}{\sinh u r} \right)   \nonumber \\
&& V_4({\bf r},u) \equiv -\frac{\sigma^a\hat{r}^a}{2} \left(u\coth ur -
\frac{1}{r} \right) 
\label{mstring}
\eeqn
It is selfdual as $B_i=F_{i4}$. The asymptotic value of
$V_4$ can be gauged away and so there is no gauge symmetry
breaking. The instanton number density per unit length along the $x^4$
direction
\beq
{\cal I} =  \frac{1}{4\pi^2} \int d^3x \Tr B_i F_{i4}
\eeq
for the above configuration is ${\cal I} = \frac{u}{2\pi}$. This is a
purely instanton string which has a long range tail.

When the scalar field takes nonzero expectation value 
asymptotically
\beq
\langle \phi \rangle = \frac{\sigma^3}{2} h 
\eeq
with positive $h>0$, the gauge symmetry is spontaneously broken to
$U(1)$ subgroup. The nontrivial scalar and $A_0$ fields of the BPS
equations (\ref{bps1}) and (\ref{laplace}) in this background are
\beq
\phi = \alpha \frac{h}{u} V_4, \;\;\; A_0 = \beta \frac{h}{u} V_4
\eeq
with $\alpha,\beta = \pm 1$ independently. Since the gauge invariant
$U(1)$ magnetic field becomes asymptotically ${2\over h}Tr (B_i{\phi})
\approx \alpha \hat{r}^i /r^2$, %with $\hat{\phi} =\hat{r}^i\sigma^i/2$ 
the total magnetic flux on the transverse three dimensions is
$4\pi\alpha $.  The instanton string now appears as magnetic or
antimonopole string depending on the sign $\alpha$.

Also the gauge invariant electric field becomes asymptotically $2\Tr (
F_{i0} \hat{\phi} )= \alpha\beta \frac{h}{u} \hat{r}^i/r^2$ and so the
electric charge density per unit length 
%defined by the conjugate
%momenta $F_{i0}/e^2$ 
becomes $2\pi \alpha\beta h/g^2 u $ and
so its magnitude is proportional to $h$ and its sign depends on the
sign $\alpha\beta$.  The linear momentum density along the $x^4$
direction is
\beq
{\cal P}_4 = - \frac{1}{g^2}\int d^3 x \Tr F_{i4} F_{i0} 
\label{4momentum}
\eeq
which is $ {\cal P}_4= -\frac{2\pi}{g^2} \beta h $.  The energy
density per unit length of this dyonic instanton string is 
\beq
{\cal H}_4 =
\frac{2\pi}{g^2}( u + \frac{h^2}{u}) \,.
\label{4energy}
\eeq

Thus our string configuration carries the instanton density, magnetic
charge, electric charge density and linear momentum density along the
string. The eigenvalues of the matrix scalar field $\phi= X_9$
describe the the deformation of $D4$ branes along the transverse ninth
direction. The zeros of the field $\phi$ are the place where two D4 branes
meet. For our string configuration the zeros lie along the line
$r=0$. Thus our solution can be interpreted as a flat charged D2 
connecting two D4 branes. As discussed in the introductory
part, this particular configuration preserves eight real supersymmetries
instead of four. This enhancement of supersymmetries occurs due to the fact
$A_4$ is proportional to $\phi$, which may be directly checked using the 
gaugino variation.

The free parameters of our dyonic-instanton-monopole string are $u$
%$h$ 
and two signs $\alpha,\beta$. Depending on $\alpha$, the magnetic
charge can be positive or negative. The sign of the electric charge
depends on $\alpha\beta$. The sign of the linear momentum depends on
$\beta$.  

To compare the flat charged D2 appearing in the above 
%configuration 
with that of the Born-Infeld description,
first we evaluate the D0 brane density from the view point of
the D2 worldvolume. Of course the D2 is extended in  $x^4$ and $x^9$ 
directions. Since D2 has a length  $(2\pi l_s^2)\, h $ in $x^9$,
the D0 brane density may be evaluated 
as\footnote{For 
simplicity, we use the convention  $2\pi l_s^2 = 2\pi g_s l_s$=1 
in the Born-Infeld description where $l_s$ and $g_s$ 
are respectively the string length scale and the string 
coupling.
In this convention, 
the D2 tension $T^{-1}_{Dp}=g_s(2\pi)^p l_s^{p+1}$ becomes 1 and the 
four dimensional Yang-Mills coupling ${ g^2}= {g_s (2\pi)^2 l_s}
=
{ 2\pi}$.}
\beq
n_{D0}=   %{{1 \over 4\pi^2}
{ %\int^{L_4}_0 d x_4\, 
{\cal I}
%\int d^3 x  \Tr B_i F_{i4}
\over
% (2\pi l_s^2) 
h %\, L_4  
}
=%{1\over  (2\pi l_s^2)} 
{u \over 2\pi h }\,.
\eeq
Hence from ${1\over 2\pi} B = n_{D0}$, we get 
\beq
B =%{1\over  (2\pi l_s^2)} 
{u \over  h }\,.
\eeq
On the other hand, the F-string density is proportional to the 
linear electric charge density along $x^4$ direction and  given by
\beq
q_{F1}={1\over 2\pi h} \int d^3 x \partial_i \Tr E_i \phi =
\Pi= {h\over u}\,,   
\eeq
reproducing $B\Pi=1$.

Using (\ref{4momentum}) and (\ref{4energy}),
the momentum and the energy density per unit area of D2 are evaluated as
\beqn
&& {\cal P}_{D2}={{\cal P}_4\over %(2\pi l^2_s) 
h}= %{1\over (2\pi l_s^2)}
{2\pi\over g^2}=1\nonumber\\
&& {\cal H}_{D2}={{\cal H}_4\over %(2\pi l^2_s) 
h}=
%{1\over (2\pi l_s^2)} 
{2\pi\over g^2}\left({u\over h}+{h\over u}\right)=
|B|+|\Pi|\,,   
\eeqn 
where we use $g^2=2\pi$ and choose $\beta=-1$.
Furthermore, the worldvolume E-field is reproduced to be
$E_{9}=1$ noting  $E_i \hat{r}_i (-dr)= E_{9} d X^9$.
Hence all the key characteristic properties of the charged D2 in the
Born-Infeld descriptions are reproduced in the above Yang-Mills
theory configuration.

One can imagine various superpositions of our configurations with
different parameters, keeping the selfduality of the four dimensional
gauge field.  Some of them would describe many parallel super charged 
D2
branes.  Finding such configurations would be daunting in general. The
simplest one would be having two  parallel  strings with identical
parameters. This solution would be independent of $x^4$ direction and
so can be obtained in principle from the BPS two monopole field
configuration.

\section{Supersymmetric monopole-antimonopole strings}

To argue that certain limit of dyonic instanton configurations are
supertubes connecting two D4 branes, we want to show that one can
superpose two parallel instanton strings of opposite magnetic
charge. This would correspond to the  supersymmetric %`super tube like' 
D2 anti-D2
branes connecting two D4 branes. Obviously such a configuration would
depend on $x^4$ coordinate nontrivially. It seems very hard to
construct such field configuration in general.  Let us try to
introduce monopole string of instanton density $u/2\pi $ of positive
electric charge and antimonopole string of instanton density $v/2\pi$
and positive electric charge. Then their fourth linear momentum
densities have opposite signs, and we choose the Lorentz frame along
$x^4$ direction so that their sum vanishes. In large separation of two
such strings, the $x^4$ dependence becomes weaker and one should be
able to identify one as magnetic string and another as antimagnetic
string. In short separation the $x^4$ dependence becomes prominent
and it becomes hard to distinguish two strings.

To find such a configuration, we regard  that the instanton density is the
sum $(u + v)/2\pi$ at large separation. Thus the instanton charge per
unit length  is
\beq \frac{1}{L} = \frac{u+v}{2\pi} \eeq
As the instanton number over the length $L $ is one and so we 
want to require that the field configuration is periodic under shift
$x^4 \rightarrow x^4+L$ up to a gauge.

Instanton configurations which are periodic in $ x^4$ direction can be
regarded as calorons, or periodic instantons.  Such  configurations
have been found with trivial symmetry breaking~\cite{harry} or with
nontrivial Wilson loop symmetry breaking~\cite{piljin,lu,kraan}. The
1/4 BPS configurations involving calorons have been also
studied~\cite{sang}. Here we start from what is known. The Wilson loop
symmetry breaking can be recasted so that the gauge field is not
periodic but $A_M(x^4+L) = U A_M(x^4) U^\dagger $ where $ U=e^{i
\frac{\sigma^3}{2} uL} $

The method to construct the solution is the Nahm's construction. The
detail of the construction is given in Refs.~\cite{lu,kraan}. The key parameters are the 
separation $D$ between two strings. In the transverse three
dimensions, the position parameters  of two strings are
\beq
{\bf x}_1 = (0,0,x_1), \;\; {\bf x}_2 = (0,0,x_2)
\eeq
with the separation parameter $D= x_2-x_1$.  The field configuration
at the position $({\bf x}, x^4)$ is expressed in terms of two
coordinates 
\beq
{\bf y}_1 = ({\bf r}-{\bf x}_1)\; , \;\;
 {\bf y}_2 = ({\bf r}-{\bf x}_2), \;\; 
\eeq
and dimensionless parameters
\beq
{\bf s}_1 = u{\bf y}_1\; , \; \;  {\bf s}_2 = v{\bf y}_2\,.
\eeq

The resulting gauge field is nontrivial superposition of two monopole
strings, 
\beq
A_\mu = -{i \partial_\mu  {\cal N}\over 2 {\cal N}^2}  
+i C_1^\dagger \partial_\mu
C_1 +i C_2^\dagger \partial_\mu C_2 + 
C^\dagger_1 V_\mu ({\bf y}_1, u) C_1 + C_2^\dagger V_\mu({\bf y}_2, v) C_2
\eeq
where the normalization factor
\beq
{\cal N} = 1 + \frac{2D}{{\cal M}} \Bigl[ N(y_1,u) (\cosh s_2 -
(\hat{y}_2)_3 \sinh s_2) + N(y_2,v) (\cosh s_1 +
(\hat{y}_1)_3 \sinh s_1 )\Bigr]
\eeq
with $N(r,u ) = \sinh ur/r  $  and ${\cal M} = 
2\left(\cosh s_1\cosh s_2+ \hat{y}_1\cdot\hat{y}_2
\sinh s_1 \sinh s_2 -  \cos \frac{2\pi x^4}{L} \right)$.
The matrices are
\beqn
&& C_1 = i\sqrt{\frac{2DN(y_1,u)}{{\cal N}} } \frac{B_1^\dagger}{{\cal
M}}\left
( e^{-\frac{1}{2} \vec{\sigma}\cdot {\bf s}_2 } Q_+  + 
 e^{\frac{1}{2} \vec{\sigma}\cdot {\bf s}_2 } Q_-   \right)
e^{-\frac{iv}{2}  \sigma^3 x^4 } \\
&& C_2 = i\sqrt{\frac{2DN(y_2,v)}{{\cal N}} } \frac{B_2^\dagger}{{\cal
M}}\left
( e^{\frac{1}{2} \vec{\sigma}\cdot {\bf s}_1 } Q_+  + 
 e^{-\frac{1}{2} \vec{\sigma}\cdot {\bf s}_1 } Q_-   \right)
e^{+\frac{iu}{2}  \sigma^3 x^4 } 
\eeqn
where
\beqn
&& B_1^\dagger  = e^{\frac{i\pi}{L}x^4} e^{-\frac{1}{2}{\sigma}\cdot
{\bf s}_1} e^{-\frac{1}{2} \sigma\cdot {\bf s}_2} 
- e^{-\frac{i\pi}{L}x^4} e^{\frac{1}{2}{\sigma}\cdot
{\bf s}_1} e^{\frac{1}{2} \sigma\cdot {\bf s}_2} \\
&& B_2^\dagger  = e^{\frac{i\pi}{L}x^4} e^{-\frac{1}{2}{\sigma}\cdot
{\bf s}_2} e^{-\frac{1}{2} \sigma\cdot {\bf s}_1} 
- e^{-\frac{i\pi}{L}x^4} e^{\frac{1}{2}{\sigma}\cdot
{\bf s}_2} e^{\frac{1}{2} \sigma\cdot {\bf s}_1} 
\eeqn
and $Q_\pm= (1\pm \sigma_3)/2$.

The solution of the scalar field equation is
\beq
\phi = \frac{h}{{\cal N}}  \frac{\sigma^3}{2}
+ \frac{H}{u}  C_1^\dagger V_4({\bf y}_1,u) C_1  
- \frac{H}{v}  C_2^\dagger V_4({\bf y}_2,v)C_2 
\eeq
where $H = \frac{ \mu D h}{1+ \mu D} $ with $\mu =
uv/(u+v)$.  Notice that this is a nonlinear superposition of the
scalar fields for each magnetic monopoles.

Some properties of the instanton part are explored in Refs.~\cite{lu,sang}.
The above configuration has  gauge singularities
at $({\bf x}=0, x^4 = n L)$ with an integer $n$. These gauge
singularities can be removed to the spatial infinity. The topological
charge can be found at the singularities.

Far from the monopole core region ($s_1,s_2\gg 1$) and $r \gg D$, 
we can neglect
exponentially small terms.  In the asymptotic region, the gauge %and scalar 
field
approaches
\beqn
 A_\mu \rightarrow  {\cal O}\left(\frac{1}{r^2}\right)\,, 
%\\
%&& \phi \rightarrow  \left(1- \frac{D}{r(1+\mu D)} \right)
%\frac{\sigma^3}{2} h 
\eeqn
and %Thus asymptotically 
the  $U(1)$   magnetic field $2\tr B_i \hat{\phi}$
falls off like $1/r^3$. This implies that the total net magnetic charge
vanishes and the system is composed of a monopole and an antimonopole.
For the electric charge densities, a straightforward evaluation leads to
\beqn
\phi \rightarrow  \left(1%- %\frac{D}{r(1+\mu D)} 
+ {q_1+q_2\over r} + {\hat{r}\cdot (q_1{\bf x}_1+ q_2 {\bf x}_2)\over r^2}
\right)\frac{\sigma^3}{2} h + {\cal O}\left(\frac{1}{r^3}\right)
\label{scalar}
\eeqn
where
\beq
q_1={\mu D\over 1+\mu D} {h\over u}\,\ \ \ 
q_2={\mu D\over 1+\mu D} {h\over v}\,. 
\label{charges}
\eeq
As for the BPS configuration $A_0=%\gamma 
\phi$, 
%with
%$\gamma=\pm 1$, 
the electric charge densities per unit length then become $q_1$
and $q_2$ for the monopole and the antimonopole strings respectively.   
%
%\beq
%q = \frac{4\pi Dh\gamma }{e^2(1+\mu D)}
%\eeq
%
These charges increase from zero to a finite value when $\mu D$ 
increases
from zero to infinity.  

The expression in (\ref{scalar}) shows that the charges $q_1$ and $q_2$
are located ${\bf x}_1$ and ${\bf x}_2$ respectively in the three 
space.  Compared with the single string case, the zeros of the field $\phi$
would not be exactly on the monopole string position. When the
separation is very large, one can calculate the $\phi$ field around
each monopole string and see how its zero gets modified. One can show
easily that the shift is at most of order $1/D^2$ 
and so essentially the zeros of the $\phi$ field lie 
along ${\bf x}_1$ and ${\bf x}_2$. This shows
that $D4$ branes are connected at the monopole and antimonopole
strings as expected for the supertube connecting two $D4$ branes. 
%The linear momentum density per unit length
%vanishes as the total magnetic charge vanishes. 

These two monopole strings carry opposite linear momentum and so their
sum vanishes.  The monopole and the antimonopole  strings carry respectively
instanton number 
densities ${\cal I}_1={u\over 2\pi}$ and 
${\cal I}_2={v\over 2\pi}$. As in the case of the single monopole string,
the worldvolume B-fields of D2 and $\overline{\rm D2}$ may be evaluated as
$B_{1}= u/h$ and $B_{2}= - v/h$ where the extra $(-)$ sign for $B_{2}$
comes from the fact that we are considering $\overline{\rm D2}$ with D0's
melted in. From the expressions of the electric charges densities,
the linear F-string densities are identified as 
%\beq
$\Pi_1=q_1$ %{\mu D\over 1+\mu D} {h\over u}\,\ \ \ 
and $\Pi_2=q_2$. %{\mu D\over 1+\mu D} {h\over v}\,. 
%\label{charges1}
%\eeq
The total energy density from the viewpoint of D2-$\overline{\rm D2}$ 
worldvolume may be checked to be  ${\cal H}_{2}
=|B_{1}|+|\Pi_1|+|B_{2}|+|\Pi_2|$. Thus the key characteristic 
properties 
of the 
supersymmetric D2 and $\overline{\rm D2}$ are reproduced except
the relation $\Pi_a=|B_{a}|^{-1} {\mu D\over 1+\mu D}$ 
for $a=1,2$. This additional 
factor $\mu D/ (1+\mu D)$ in 
the F-string densities is resulted from the effect of the D4 branes 
to the supersymmetric D2 and $\overline{\rm D2}$ system. For the large
 separation, $\mu D/ (1+\mu D)$ approaches one and the effect disappears
as expected.

A few comments are in order.
The gauge field and the scalar field is not invariant under $x^4
\rightarrow x^4+L$. It transforms as
\beq
(A_\mu, \phi)(x^4+L) =  e^{-i \frac{\sigma^3}{2} uL} (A_\mu,
\phi)(x^4) 
e^{i \frac{\sigma^3}{2} u L} 
\eeq
We can make a nonsingle-valued gauge transformation $U=e^{-\frac{i}{2}
\sigma^3 u x^4} $ to get  periodic gauge and scalar fields,  in which
case the $A_4$ field would have nontrivial expectation value
\beq
\langle A_4 \rangle = \frac{\sigma^3}{2} u\,. 
\eeq
In the compact $x^4$ case, this expectation value becomes the
 Wilson loop which cannot be gauged away~\cite{lu,kraan}.

%When, say, $v=0$, the electric charge is carried mostly
%carried by the second monopole string which corresponds to the
%massless monopole string. Thus, in the limit where $\mu D$ goes to
%infinity, we end up with only neutral instanton string in the broken
%phase, which is impossible as we show remain in the BPS limit.  The
%way out is to notice that the above formula is correct only up to
%order $1/D$ correction which becomes important when $v=0$ and $D$ is
%large.

In the limit either $u$ or $v$ vanishes, the above solution becomes
the conventional caloron solution found before~\cite{harry}.  A single
caloron is made of a massive monopole and massless monopole when $D$
is large.  The electric charge would increase linearly with the string
separation $D$.

%Let us now examine the above solution more closely. 
%When $D>> 1/u. 1/v$, two monopole strings are well separated. Near
%each monopole string, the gauge and scalar field approach a gauge
%transformation of those for a single monopole string.
%After the gauge transformation, they  are selfdual. The electric
%charge density and linear momentum density carried by the first
%monopole string are
%
%\beqn
%&& q_1 = \frac{4\pi  Dh\gamma}{e^2(1+\mu D)} \frac{v}{u+v} \\
%&& p_1 = \frac{4\pi \mu D h}{e^2(1+ \mu D)} 
%\eeqn
%
%to order $1/D$ correction. 
%Those carried by the second monopole string is
%
%\beqn
%&& q_2 = \frac{4\pi  Dh\gamma}{e^2(1+\mu D)} \frac{u}{u+v} \\
%&& p_2 = -\frac{4\pi \mu D h}{e^2(1+ \mu D)} 
%\eeqn
%

In short distance separation with $D\ll L$, the caloron solution
becomes basically a periodic array of well separated instantons.  The
zeros of the $\phi$ field are expected to change from two separated
line shape to a periodic array of isolated regions of zeros. When
$D=0$, the instantons get very small and so the positions of zeros would
be a periodic array of the instanton positions.  We do not know how
this transition occurs exactly.

\section{Conclusion}

We have studied the various aspects of supertubes connecting D4
branes. Especially we found and explored the dyonic instanton
configuration which corresponds to dyonic monopole and antimonopole
strings. They are identified as the supersymmetric D2-$\overline{\rm
D2}$ connecting two D4 branes. Compared to the case of supersymmetric
D2-$\overline{\rm D2}$ without D4 branes, the effect of D4 branes
appears in the F-string densities by the factor $\mu D/(1+\mu D)$
depending on the separation $D$ of D2 and $\overline{\rm D2}$.

The dyonic instantons found by the 't Hooft ansatz have only isolated
zeros and so could not be identified with supertubes connecting two D4
branes. For these solutions, the angular momentum of the selfdual
dyonic instanton is antiselfdual. This is not true for our single
monopole string, in which case the angular momentum with respect to
any reference point has both selfdual and antiselfdual component.
Clearly our solutions are outside the 't Hooft ansatz family.

While our examples for supertubes have infinite instanton number, we 
wonder a possibility that supertubes may appear even with finite
instanton number for generic dyonic configurations beyond the 't Hooft
ansatz.  

Finally, the study of  moduli dynamics of the charged 
monopole string would be quite 
interesting including the interactions between the charged 
monopole and antimonopole strings.

\subsection*{Acknowledgement}
This work is supported in part by KOSEF 1998 interdisciplinary
research grant 98-07-02-07-01-5. We acknowledge interesting discussions
with N. Ohta and P. Yi.

%\newpage


\begin{thebibliography}{99}

\bibitem{mateos}
D.~Mateos and P.~K.~Townsend,
%``Supertubes,''
Phys.\ Rev.\ Lett.\  {\bf 87}, 011602 (2001)
[arXiv:hep-th/0103030].
%%CITATION = HEP-TH 0103030;%%


%\cite{Bak:2001kq}
\bibitem{klee}
D.~Bak and K.~Lee,
%``Noncommutative supersymmetric tubes,''
Phys.\ Lett.\ B {\bf 509}, 168 (2001)
[arXiv:hep-th/0103148].
%%CITATION = HEP-TH 0103148;%%

%\cite{Bak:2001gm}
\bibitem{swkim}
D.~Bak and S.~W.~Kim,
%``Junctions of supersymmetric tubes,''
Nucl.\ Phys.\ B {\bf 622}, 95 (2002)
[arXiv:hep-th/0108207].
%%CITATION = HEP-TH 0108207;%%


%\cite{Bak:2001xx}
\bibitem{karch}
D.~Bak and A.~Karch,
%``Supersymmetric brane-antibrane configurations,''
Nucl.\ Phys.\ B {\bf 626}, 165 (2002)
[arXiv:hep-th/0110039].
%%CITATION = HEP-TH 0110039;%


%\cite{Bak:2001tt}
\bibitem{ohta}
D.~Bak and N.~Ohta,
``Supersymmetric D2 anti-D2 strings,''
Phys.\ Lett.\ B {\bf 527}, 131 (2002)
[arXiv:hep-th/0112034].
%%CITATION = HEP-TH 0112034;%%







%\cite{Mateos:2001pi}
\bibitem{ng}
D.~Mateos, S.~Ng and P.~K.~Townsend,
%``Tachyons, Supertubes and Brane/Antibrane Systems,''
JHEP {\bf 0203}, 016 (2002)
[arXiv:hep-th/0112054].
%%CITATION = HEP-TH 0112054;%%



%\cite{Kruczenski:2002mn}
\bibitem{kruczenski}
M.~Kruczenski, R.~C.~Myers, A.~W.~Peet and D.~J.~Winters,
%``Aspects of supertubes,''
JHEP {\bf 0205}, 017 (2002)
[arXiv:hep-th/0204103].
%%CITATION = HEP-TH 0204103;%%




%\cite{Bak:2002wy}
\bibitem{jabbari}
D.~Bak, N.~Ohta and M.~M.~Sheikh-Jabbari,
``Supersymmetric brane anti-brane systems: 
%Matrix model description,  stability and decoupling limits,''
arXiv:hep-th/0205265.
%%CITATION = HEP-TH 0205265;%%


%\cite{Lugo:2002wr}
\bibitem{lugo}
A.~R.~Lugo,
``On supersymmetric $Dp - \bar D p$ brane solutions,''
arXiv:hep-th/0206041.
%%CITATION = HEP-TH 0206041;%%



\bibitem{cho}
J.H.~Cho and P.~Oh,
%``Super D-helix,''
Phys.\ Rev.\ D {\bf 64}, 106010 (2001)
[arXiv:hep-th/0105095].


%\cite{Emparan:2001ux}
\bibitem{emparan}
R.~Emparan, D.~Mateos and P.~K.~Townsend,
%``Supergravity supertubes,''
JHEP {\bf 0107}, 011 (2001)
[arXiv:hep-th/0106012].
%%CITATION = HEP-TH 0106012;%%

%\cite{Mateos:2002pi}
\bibitem{townsend3}
D.~Mateos, S.~Ng and P.~K.~Townsend,
``Supercurves,'' [arXiv:hep-th/0204062].

\bibitem{ohta2}
Y.~Hyakutake and N.~Ohta,
``Supertubes and Supercurves from M-Ribbons,''
[arXiv:hep-th/0204161].





\bibitem{adhm}
M.~F.~Atiyah, N.~J.~Hitchin, V.~G.~Drinfeld and Y.~I.~Manin,
%``Construction Of Instantons,''
Phys.\ Lett.\ A {\bf 65} (1978) 185.
%%CITATION = PHLTA,A65,185;%%




%\cite{Lambert:1999ua}
\bibitem{lambert}
N.~D.~Lambert and D.~Tong,
%``Dyonic instantons in five-dimensional gauge theories,''
Phys.\ Lett.\ B {\bf 462}, 89 (1999)
[arXiv:hep-th/9907014].
%%CITATION = HEP-TH 9907014;%%


%

%\cite{Zamaklar:2000tc}
\bibitem{zamaklar}
M.~Zamaklar,
%``Geometry of the nonabelian DBI dyonic instanton,''
Phys.\ Lett.\ B {\bf 493}, 411 (2000)
[arXiv:hep-th/0006090].
%%CITATION = HEP-TH 0006090;%%


%\cite{Eyras:2000dg}
\bibitem{eyras}
E.~Eyras, P.~K.~Townsend and M.~Zamaklar,
%``The heterotic dyonic instanton,''
JHEP {\bf 0105}, 046 (2001)
[arXiv:hep-th/0012016].
%%CITATION = HEP-TH 0012016;%%

\bibitem{sang}
K. Lee and S.H. Yi, 
``1/4 BPS dyonic calorons,''
[arXiv:hep-th/0205274].
%%CITATION = HEP-TH 0205274;%%

%\cite{Harrington:1978ve}
\bibitem{harry}
B.~J.~Harrington and H.~K.~Shepard,
%``Periodic Euclidean Solutions And The Finite Temperature Yang-Mills Gas,''
Phys.\ Rev.\ D {\bf 17}, 2122 (1978);
%%CITATION = PHRVA,D17,2122;%%
%\cite{Rossi:qe}
%\bibitem{rossi}
P.~Rossi,
%``Propagation Functions In The Field Of A Monopole,''
Nucl.\ Phys.\ B {\bf 149}, 170 (1979).
%%CITATION = NUPHA,B149,170;%%

%\cite{Lee:1997vp}
\bibitem{piljin}
K.~Lee and P.~Yi,
%``Monopoles and instantons on partially compactified D-branes,''
Phys.\ Rev.\ D {\bf 56}, 3711 (1997)
[arXiv:hep-th/9702107].
%%CITATION = HEP-TH 9702107;%%


%\cite{Lee:1998bb}
\bibitem{lu}
K.~Lee and C.~Lu,
%``SU(2) calorons and magnetic monopoles,''
Phys.\ Rev.\ D {\bf 58}, 025011 (1998)
[arXiv:hep-th/9802108].
%%CITATION = HEP-TH 9802108;%%



%\cite{Kraan:1998sn}
\bibitem{kraan}
T.~C.~Kraan and P.~van Baal,
%``Monopole constituents inside SU(n) calorons,''
Phys.\ Lett.\ B {\bf 435}, 389 (1998)
[arXiv:hep-th/9806034].
%%CITATION = HEP-TH 9806034;%%



\end{thebibliography}
\end{document}